# Two-photon dual-comb LiDAR imaging

ALEXANDER J. M. NELMES,[1,†] SIMON FLETCHER[2], ANDREW LONGSTAFF[2], JAKE M. CHARSLEY,[1] HOLLIE WRIGHT,[1,†] AND DERRYCK T. REID[1,*]

[1]*Scottish Universities Physics Alliance, Institute of Photonics and Quantum Sciences, School of Engineering and Physical Sciences, Heriot-Watt University, Edinburgh, EH14 4AS, UK*
[2]*Centre for Precision Technologies (CPT), University of Huddersfield, Queensgate campus, Huddersfield, HD1 3DH, UK*
[†]*These authors contributed equally to this work.*
* D.T.Reid@hw.ac.uk

**Conventional LiDAR uses time-of-flight data from laser pulses scanned across a scene to provide accurate multi-meter-scale three-dimensional models at cm precision, limited by the tens-of-picoseconds precision of time-tagging electronics. Here, by using two-photon dual-comb ranging, we introduce an analog of LiDAR imaging using the time-of-flight of sub-picosecond laser pulses to render cm-scale point-cloud datasets with µm precision. Using only free-running femtosecond lasers, the technique combines absolute accuracy with near-interferometric precision, is applicable to discontinuous surfaces with poor optical quality, and provides a stand-off range exceeding that of other optical metrologies. We demonstrate imaging of an aluminum test object and assess its accuracy by comparing our results with those from a touch-probe coordinate measurement machine. At a stand-off distance of 40 cm, we obtain ranging accuracies of 9 µm–38 µm, and precisions averaging to 1.0 µm after 500 ms.**

Dual-comb ranging [1,2] is a non-mechanical dimensional metrology technique that is distinguished by its unique combination of features, including: the ability to measure absolute distance unambiguously over meter scales; sub-µm precision for optical-quality targets [3]; and self-calibration, relying on only one experimental parameter, the probe-comb pulse repetition rate. Three-dimensional surface profiling has been reported using comb-calibrated frequency modulated continuous-wave laser ranging [4] and, recently, by direct coherent dual-comb ranging [5], but its precision is limited in each approach to 5 µm–10 µm by speckle noise from diffusely scattering targets [6]. The complementary approach of two-photon dual-comb LiDAR eliminates speckle by replacing coherent gating with non-interferometric two-photon cross correlation [7,8], while simultaneously avoiding the need for complex phase-stabilized dual-comb lasers and ultra-high-speed GS/s fringe-resolving data acquisition [9]. Instead of generating an interferogram for each sampling point, pairs of probe and local-oscillator (LO) pulses create a single electrical pulse, which is time-stamped to achieve (after averaging) 100-nm precisions for specular targets. Here, we introduce an exact dual-comb analog of LiDAR imaging that exploits this efficient and scalable data-acquisition approach to achieve point-cloud imaging of a diffusely scattering test object and single-point precisions of 1 µm. We compare the precision and accuracy obtained with measurements made using an industry-standard coordinate measurement machine (CMM), finding agreement to within 10 µm accuracy.

The fully fiberized dual-comb imaging system (Fig. 1(a)) employed two free-running 78 MHz Er:fiber lasers, operating with pulse durations of 500 fs and with a repetition-frequency difference of 1 kHz. The probe-comb pulses simultaneously interrogated a reference reflector and the test object in a common transmit-receive (TX/RX) geometry, with amplification of the returned pulses in an Er:fiber amplifier (EDFA; Thorlabs, EDFA1000P) allowing the recovery of high-precision ranging from diffusely-scattering metal surfaces, even when the optical power was attenuated to <1% after probing the target [10]. After amplification, the probe pulses returned from the reference and target were combined with the LO pulses in a fiber combiner, then cross-correlated in a laser diode (LD; Luminent MRLDFC010) operating as a two-photon absorption detector [11,12], resulting in a sequence of alternating LO-reference and LO-target cross-correlations. A 30 MHz low-pass filter (Mini Circuits, BLP-30+) was used to suppress frequencies at the 78 MHz laser repetition rate, after which the cross-correlations were amplified by 200× in a high-speed current amplifier (Femto, HCA-20M-100K-C). Finally, a constant-fraction discriminator (CFD; FlimLabs, MSFL0006) was used to produce CMOS-compatible pulses with rising edges aligned exactly to the peak of each cross-correlation. The CFD increases the timing resolution by avoiding the edge jitter that would occur if a comparator or other thresholding device were used to condition cross-correlation pulses of varying voltage levels [13]. These electrical pulses were time-stamped to 1.67 ns precision using a Teensy 4.0 microcontroller to provide a continuous stream of 32-bit integers representing the time intervals between consecutive target and reference pulses. Further details of the electronic conditioning and the comb sources are given in [10].

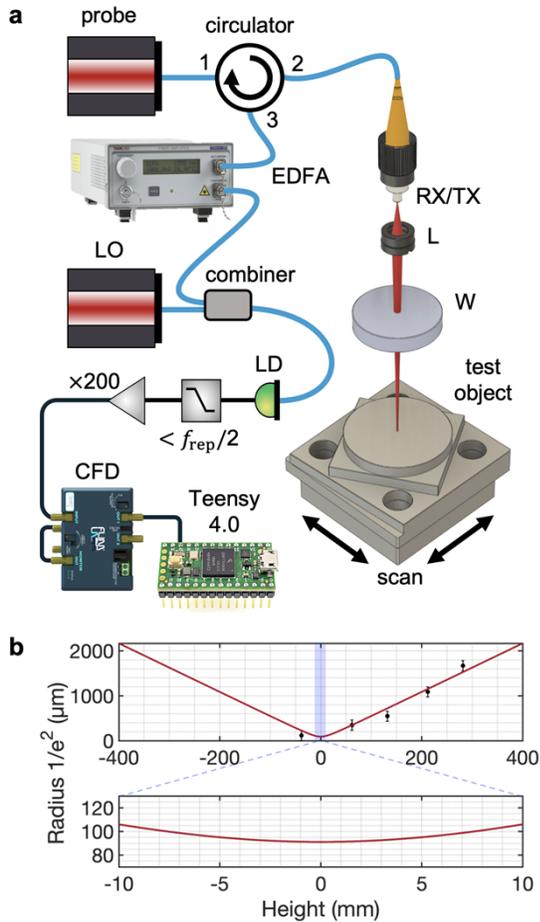

Fig. 1. (a) Two-photon dual-comb LiDAR system. RX/TX: receive / transmit aperture; L, lens; W, reference wedge; EDFA: Er-doped fibre amplifier; LD: laser diode; CFD: constant fraction discriminator. Blue lines indicate optical fiber. Black lines indicate electrical connections. (b) Measured (symbols) and fitted (solid red lines) beam radius, where the height is measured from the mid-point of the target-object surfaces.

A precision motorized stage (ASI Imaging MS-2000) supported the test object, allowing it to be translated under the probe beam to programmed ($x,y$) coordinates. The probe beam from an FC/PC connector was collimated to create a loose focus in the center of the object. The aim of this alignment was to create a beam with a consistent diameter throughout the extent of the test object, corresponding to a confocal parameter of a few cm, and an aspheric lens with a focal length of 18.4 mm (Fig. 1(a), L; Thorlabs, A280TM-C) was selected for this purpose. Knife-edge measurements characterizing the beam (Fig. 1(b)) indicated a focal spot radius of 91 µm, corresponding to a confocal parameter of 34 mm. The test object surfaces had a height range of 19.2 mm, over which the beam diameter change was less than ±10% (see Fig. 1(b), lower panel).

A glass wedge (Fig. 1(a), W) was situated 340 mm above the test object and served as the reference reflector. Both the reference and test object were localized within the same non-ambiguity range of the measurement, which for the 78 MHz laser repetition rate was 1.9 m. Adjusting the angle of the wedge allowed the power returned from the reference to be made similar to that from the test object, avoiding saturation of the signal conditioning electronics. When sufficient light is returned from the test object, the sequence of cross-correlations alternates strictly between reference and target signals, resulting in time-interval data which toggle between high and low values. Where a region of low backscattering fails to return a target signal, this alternating sequence is broken, and a simple software filter was used to detect such an error and remove these data. With the addition of only the probe-comb pulse repetition frequency, these timing interval data are sufficient to extract range information at sub-µm precision [7] and with a minimal data burden in comparison to interferometric sampling methods.

The test object (Fig. 2) was a computer numerical control (CNC)-milled aluminum design produced and characterized to determine the accuracy of a milling machine, scaled down for compatibility with the scanning method. With a volume of 50×50×30 mm$^3$, it contained several distinct surfaces: circle, diamond, square, ledges and shoulders. Machining lines and a text engraving are visible.

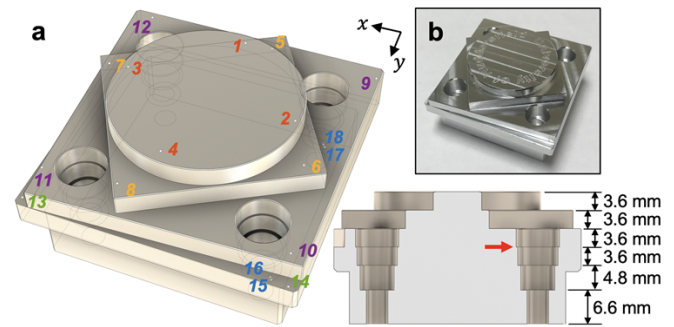

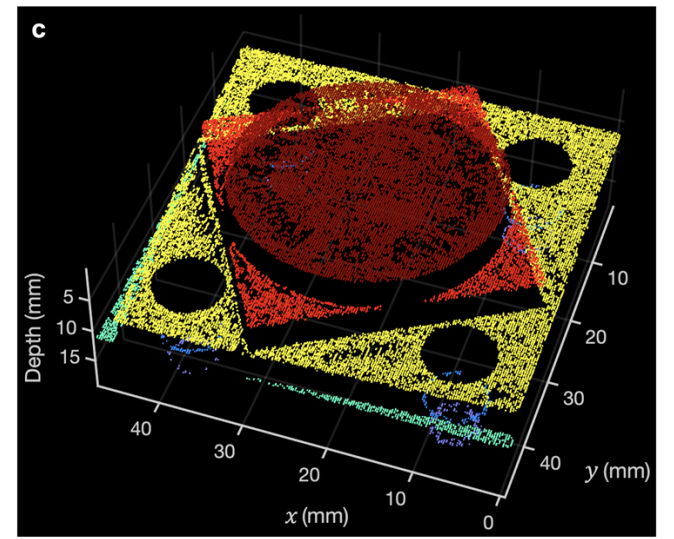

**Fig. 2.** (a) Test object CAD model, showing surface heights in cross-section, and (b) physical object. (c) Two-photon dual-comb LiDAR point cloud, showing depth-resolved imaging, with six planes visible in an animated 360° rotation of the point cloud data (see Visualization 1).

The LiDAR point-cloud image is shown in Fig. 2(c), and contains range data across a 200 × 250 grid of sampling points laterally spaced at 200 μm–250 μm intervals. A clustering algorithm was employed to automatically partition the sampled points to the distinct target object surfaces for color-coding the point cloud. Initially, an 80-element depth array was created with a spacing of 250 μm between the extremes of the lengths from 0 mm to 20 mm, sufficient for assigning each depth value to a unique surface while reducing the computational cost. Each valid lateral pixel (discounting NaN returns) was assigned a sparse array with a single non-zero value of one at the nearest depth position to that extracted. This conditioning enabled principal component analysis (PCA) to be applied to reduce the dimensionality from 80 to 10 as well as lower the cost for the subsequent $k$-means clustering. We used $k$-means clustering to partition the data into nine clusters, yielding eight depth clusters and a nineth noise cluster, representing erroneous depth values associated with poor reflectivity or edges. Two pairs of valid clusters shared the same depths, leading to a final result of six unique surface clusters, which we represent in Fig. 2(c) each with a unique color.

With the exception of only the shallowest hole depth that had only a 300 μm shoulder (Fig. 2(a), red arrow), and which the lateral sampling resolution was insufficient to resolve, all of the principal plane depths in the test object are resolved in the point-cloud image. The text engraving is partially resolved on the top surface. Gaps in each surface indicate sampling points where target returns were insufficient to trigger a measurement. In the accompanying online visualization (Visualization 1), all the main features of the point cloud data are visible.

To understand the accuracy of the measurement, surface depths extracted from the dual-comb LiDAR image were compared with the nominal feature depths extracted from the CAD model of the test object (Fig. 3, red symbols) and with measurements of the same object made using a Zeiss Calypso coordinate measurement machine (CMM; Fig. 3, blue symbols). The CMM measurement used a mechanical probe which was too large to access surface features deeper than 10.8 mm, so we restrict our comparison with the CMM to the first three surfaces.

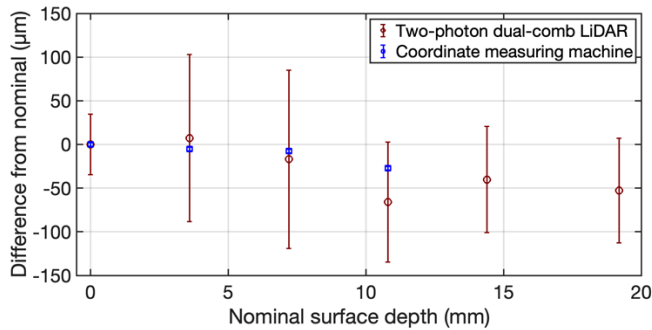

**Fig. 3.** Comparison of nominal depths with those inferred from two-photon dual-comb LiDAR image measurements (red) and from Zeiss Calypso CMM measurements (blue).

The error bars in Fig. 3 indicate the standard deviation of each of the dual-comb measurements extracted from the data shown in Fig. 2(c) and the uncertainty of the CMM (2 μm).

The top surface (circle feature) is defined as a surface depth of 0 mm in each case. For the first three surfaces, the height differences between the two-photon dual-comb LiDAR and CMM measurements are 12 μm, 9 μm and 38 μm. The poorer accuracy of the third surface may be associated with the smaller number of points available for ranging.

The point precision of the dual-comb metrology was measured separately to the imaging data by probing the 18 specific points on the test object shown in Fig. 2(a). These points were chosen to correspond to locations on the different planes in the object. The measurement precision was determined from an Allan deviation analysis (Fig. 4) of each of the 18 single-point measurements. The depth value associated with each point is the mean of up to 1000 samples acquired in less than one second. Even with the free-running femtosecond lasers used in the measurement, the Allan deviation after 500 ms falls to below 1 μm for the majority of points, and shows a $\tau^{-1/2}$ dependence, indicating that increasing the measurement dwell time would lead to even better precision. The performance is consistent with earlier observations [7,10] which show a modest reduction in precision when using diffusely scattering targets, in comparison to specular reflecting targets.

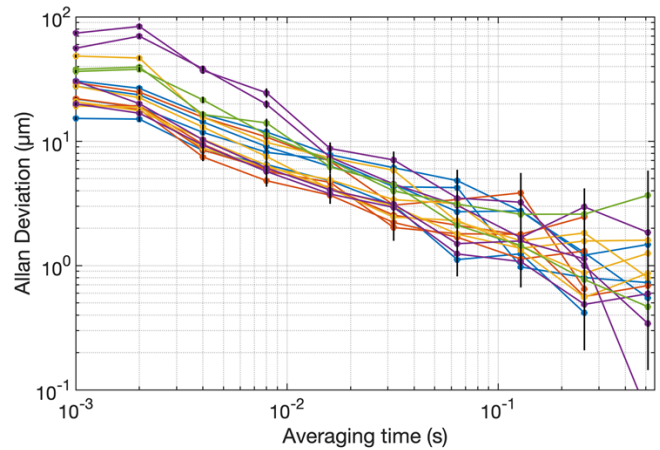

**Fig. 4.** Allan deviations for depth-measurement time-series, showing averaging to <1 μm precision for most probe points.

The absolute and comparative measurements presented here show that two-photon dual-comb LiDAR is competitive with contemporary non-contact approaches for part surface measurements in a manufacturing context, for example laser triangulation or 3-D photogrammetry, which have an accuracy between 10 μm and 100 μm and system-dependent resolutions as low as 1 μm. The agreement with independent CMM measurements, which can be considered a test of the absolute measurement accuracy, was between 9 μm and 38 μm for three surfaces. Importantly, all measurements were achieved at a stand-off range of around 40 cm, unlike those from a CMM which uses touch probes to sample an object's surface at discrete points. Two-photon dual-comb LiDAR is subject to fewer restrictions on proximity and access than a physical probe, and the imaging in Fig. 2 illustrates its ability to observe and measure high aspect ratio features, such as the holes at depths of 14.4 mm and 19.2 mm. Its high stand-off distance also compares favorably to other optical techniques such as chromatic confocal and white light interferometry, which are often limited by the use high numerical aperture objective lenses.

While using sample scanning avoids aberrations associated with beam scanning and allows repeatable single-point range measurements, it cannot deliver high speed LiDAR imaging. However, two-photon dual-comb LiDAR is in principle well suited to much higher scan rates due to its very low data processing burden and inherent ability to continuously stream ranging data. Galvanometer-mirror scanning using a telecentric objective lens geometry would allow images with a similar quality to Fig. 2(c) to be acquired in seconds to minutes, limited only by the pulse repetition frequency difference of the LO and probe combs. Combining this with precision sample positioning would allow rapid survey imaging to be combined with sub-µm precision measurements at fiducial locations, for example to enable the verification of specific feature dimensions. Aliasing limitations and related considerations for two-photon LiDAR [7] restrict the sampling rate to $f_{\rm rep}^2/2\Delta\nu$, where $\Delta\nu$ is the optical bandwidth. By increasing $f_{\rm rep}$ and / or decreasing $\Delta\nu$, higher sampling rates compatible with near-video-frame-rate acquisition should be possible. In parallel work we have used two Er,Yb:glass lasers operating at 540 MHz to achieve continuously streamed two-photon dual-comb LiDAR measurements at 10 kHz–20 kHz, with precisions averaging to below 1 µm in only 10 ms [14]. This system would in principle permit a 200 × 200 resolution image to be recorded in only two seconds, offering a practical inspection technique for use in manufacturing environments.

**Funding.** Engineering and Physical Sciences Research Council (EP/Z53285X/1). Royal Academy of Engineering (RCSRF2223-1678). Renishaw plc.

**Disclosures.** The authors declare no conflicts of interest.

**Data Availability.** Data underlying the results presented in this paper are not publicly available at this time but may be obtained from the authors upon reasonable request.